\documentclass[fleqn,twoside]{article}
\usepackage{espcrc2}

\usepackage{graphicx}
\usepackage{epsfig}
\usepackage[figuresright]{rotating}

\newcommand{\AmS}{{\protect\the\textfont2
  A\kern-.1667em\lower.5ex\hbox{M}\kern-.125emS}}

\newcommand{\eexp}{{\rm e}}

\newcommand{\Tr}{{\rm Tr}}
\newcommand{\vev}[1]{ \langle #1 \rangle }

\title{The Factorization Method for Monte Carlo Simulations of Systems
With a Complex Action}

\author{J. Ambj\o rn\address[NBI]{The Niels Bohr Institute,
                 Blegdamsvej 17, DK-2100 Copenhagen \O, Denmark}, 
        K. N. Anagnostopoulos\address[UOC]{Department of Physics, 
           University of Crete, P.O. Box 2208, GR-71003 Heraklion, Greece},
        J. Nishimura\address[KEK]{Theory Division, KEK, Tsukuba,
                 Ibaraki 305--0801, Japan}
        and
        J.J.M.  Verbaarschot\address[SUNY]{ Department of Physics
                 and Astronomy, SUNY, Stony Brook, NY 11794, USA}}
       
\begin{document}

\begin{abstract}
We propose a method for Monte Carlo simulations of systems with a complex
action. The method has the advantages of being in principle applicable
to any such system and provides a solution to the overlap problem. In
some cases, like in the IKKT matrix model, a finite size scaling
extrapolation can provide results for systems whose size would make it
prohibitive to simulate directly.

%By constructing distribution
%functions of observables it is possible to clearly see the effect of the
%imaginary part of the action.  

\vspace{1pc}
\end{abstract}

\maketitle

\section{INTRODUCTION}
There exist many interesting systems in high energy physics whose
action contains an imaginary part, such as QCD at finite baryon
density, Chern-Simons theories, systems with topological terms (like
the $\theta$-term in QCD) and systems with chiral fermions. This
imposes a severe technical problem in the simulations, requiring an
exponentially large amount of data for statistically significant
measurements as the system size is increased or the critical point is
approached. Furthermore, the overlap problem appears when standard
reweighting techniques are applied in such systems and it becomes
exponentially hard with system size to visit the relevant part of the
configuration space. In \cite{sign} it was proposed to take advantage
of a factorization property of the distribution functions of the
observables one is interested to measure. This approach can in
principle be applied to {\it any} system and it eliminates the overlap
problem completely. In some cases it is possible to use finite size
scaling to extrapolate successfully to large system sizes where it
would have been impossible to measure oscillating factors
directly. The method has been applied successfully in matrix models of
non perturbative string theory (IKKT) \cite{sign}, random matrix theory
of finite density QCD (RMT) \cite{rmt} as well as the 2d ${\rm CP}^3$
model, the 1d antiferromagnetic model with imaginary $B$ and the 2d
compact U(1) with topological charge \cite{Azcoiti}.

In this paper we present our results \cite{sign,rmt} for IKKT and
RMT. In the first case we study the space--time dimensionality hoping
to {\it dynamically} recover our 4d space--time and in the second to
test the factorization method against known analytical results.
In all cases we will be dealing with a system defined by a partition
function $Z=\int d{A}\, \eexp^{-S_{ 0}}\, \eexp^{i\Gamma}$ and the
corresponding phase quenched model $Z_0=\int d{ A}\, \eexp^{-S_{ 0}}$
where $S=S_0-i\Gamma$ is the action of the system with its real and
imaginary parts. $A$ represents collectively the degrees of freedom
of the model and in our case it corresponds to a set of $N\times N$
matrices. In case we are interested in measuring some observable
${\cal O}$, we consider the distribution functions $\rho_{\cal
O}(x)=\vev{\delta(x-{\cal O})}$ and $\rho^{(0)}_{\cal
O}(x)=\vev{\delta(x-{\cal O})}_0$, where $\vev{\ldots}_0$ refers to $Z_0$. 
Then we define the fiducial system 
$Z_{{\cal O},x}=\int d{A}\, \eexp^{-S_{ 0}} \delta(x-{\cal O})$,
the weight factor $w_{\cal O}(x)=\vev{\eexp^{i\Gamma}}_{{\cal O},x}$ and the
distribution 
$\rho_{\cal O}(x)$ factorizes
%\begin{equation}
%\label{1}
$\rho_{\cal O}(x)=\frac{1}{C}\,\rho_{\cal O}(x)\, w_{\cal O}(x)$ %\, ,
%\end{equation}
where $C=\vev{\eexp^{i\Gamma}}_0$. Then $\vev{{\cal
O}}=\frac{1}{C}\int_{-\infty}^{\infty}dx\,x\,\rho^{(0)}_{\cal
O}(x)\,w_{\cal O}(x)$. The $\delta$--function constraint is
implemented in our simulations by considering the system $Z_{{\cal
O},V}=\int dA \,\eexp^{-S_0}\,\eexp^{V({\cal O})}$ where
$V(z)=\frac{1}{2}\gamma(z-\xi)^2$ and $\gamma, \xi$ are parameters
which control the constraining of the simulation. 
The results are insensitive to the choice of
$\gamma$ as long as it is large enough. Then we have that $w_{\cal
O}(x=\vev{{\cal O}}_{i,V})=\vev{\eexp^{i\Gamma}}_{i,V}$. The
distribution of $\cal O$ in $Z_{i,V}$ has a peak $\bar x$ and the
quantity $V'(\bar x)$ is the value of $f^{(0)}_{\cal
O}(x)=\frac{d}{dx}\ln \rho^{(0)}_{\cal O}(x)$ at $x=\bar x$. The
function $\rho^{(0)}_{\cal O}(x)$ can be obtained by integrating an
analytic function to which  we fit the $f^{(0)}_{\cal O}(x)$ data points.

By applying this method we force the system to sample configurations
which give the essential contributions to $\vev{{\cal O}}$, something
that would be increasingly difficult with system size in the phase
quenched model, eliminating this way the overlap problem. This already
allows us get close to the thermodynamic limit with modest computer
resources. Furthermore we obtain direct knowledge of
$w_{\cal O}(x)$ and $\rho_{\cal O}(x)$ which allows us to understand
the effect of $\Gamma$. This is important for understanding the
properties of the system when $\Gamma$ plays a crucial role. Using the
generic scaling properties of the weight factor $w_i(x)$, one may
extrapolate the results obtained by direct Monte Carlo evaluations to
larger system size.  Such an extrapolation is expected to be
particularly useful in cases where the distribution function turns out
to be positive definite.  In those cases we can actually even {\em
avoid} using the reweighting formula by reducing the question of
obtaining the expectation value to that of finding the minimum of the
free energy, which is (minus) the {\it log} of the distribution
function.  Here, the error in obtaining the scaling function
propagates to the final result without significant magnifications.
Therefore, the extrapolation can be a powerful tool to probe the
thermodynamic limit from the accessible system size.

\section{NON--PERTURBATIVE STRING THEORY}
The so--called IKKT matrix model \cite{ikkt} has been conjectured to be
a definition for non--perturbative string theory. A particularly
interesting feature of the model is that the eigenvalues of the
bosonic matrices $A_\mu$ generate space-time {\it dynamically}. In our
case we work with a reduced model where $\mu=1,\ldots,6$. The
observables that we study are the normalized eigenvalues
$\lambda_1/\vev{\lambda_1}_0>\ldots>\lambda_6/\vev{\lambda_6}_0$ of
the space--time ``moment of inertia'' 
$T_{\mu\nu}=\frac{1}{N}\Tr A_\mu A_\nu$. An interesting scenario to
investigate would be that the O(6) symmetry is spontaneously broken
with some of the eigenvalues (possibly 4) grow large and the rest
remain small, providing a mechanism for dynamical compactification of
extra dimensions.
%%%%%%%%% FIGURE 
\vskip 6mm
  \begin{center}
 {\epsfig{file=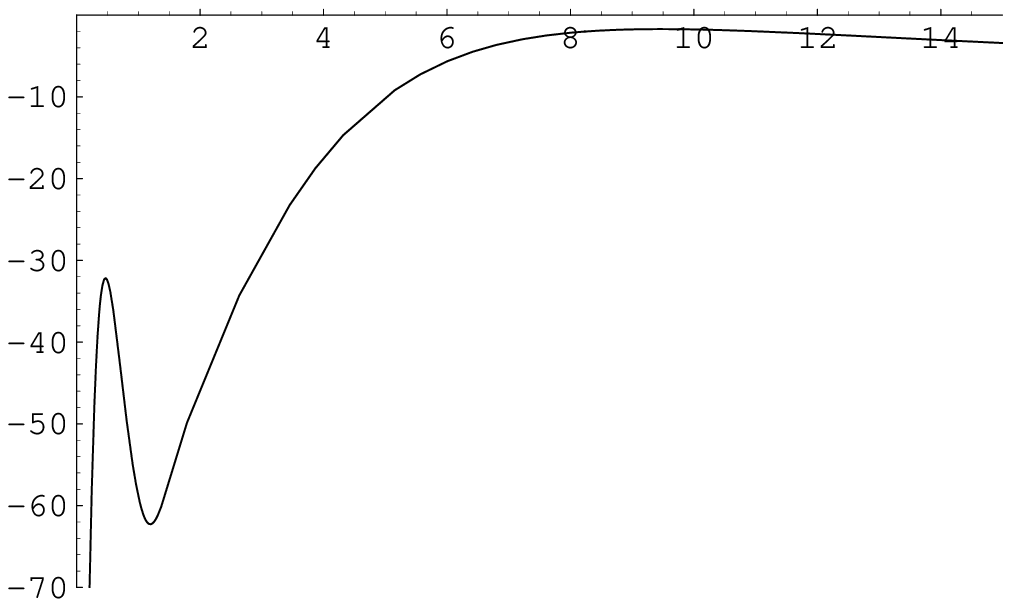,width=75mm}}
\end{center}
{\bf Fig. 1}: {\it $\log\rho_5(x)$ for N=128, n = 16.}
\vskip 6mm
%%%%%%%%% END FIGURE

The IKKT model is a good example where finite size scaling for the
oscillating factor $w_i(x)$ works well (the index $i$ corresponds to
$\lambda_i/\vev{\lambda_i}_0$). For $i>1$ we find fast convergence to
a scaling function $\Phi_i(x)$, where $w_i(x)=\exp\{N^2\Phi_i(x)\}$
\cite{sign}. We compute $\Phi_i(x)$ for small matrix size $n\le 20$
and the phase quenched distribution function $\rho_i^{(0)}(x)$ for
larger size $N$. Then the factorization formula can be used in order
to compute $\rho_i(x)$. Note that in the
computation of $\vev{\lambda_i}$ the errors do not propagate
exponentially with system size since $\vev{\lambda_i}$ can be
determined from the minimum of the ``free energy'' $F_i(x) =
-\frac{1}{N^2}\log\rho_i(x)$.  In Fig. 1 we show our results for
$\rho_5(x)$ for $n=16$ and $N=128$.  A double peak structure of
$\rho_5(x)$ is evident and it is expected that the peak for small $x$
will increase with system size \cite{sign}. One hopes that this peak
will be dominant for $i=5,6$ and that the large $x$ peak will be
dominant for $i=4,3,2,1$ and this will realize the SSB scenario. Note
the heavy suppression of the $x=1$ region caused by $\Gamma$ which is
the peak of the phase quenched model distribution of $\lambda_i$.

%%%%%%%%% FIGURE 
\vskip 6mm
  \begin{center}
 {\epsfig{file=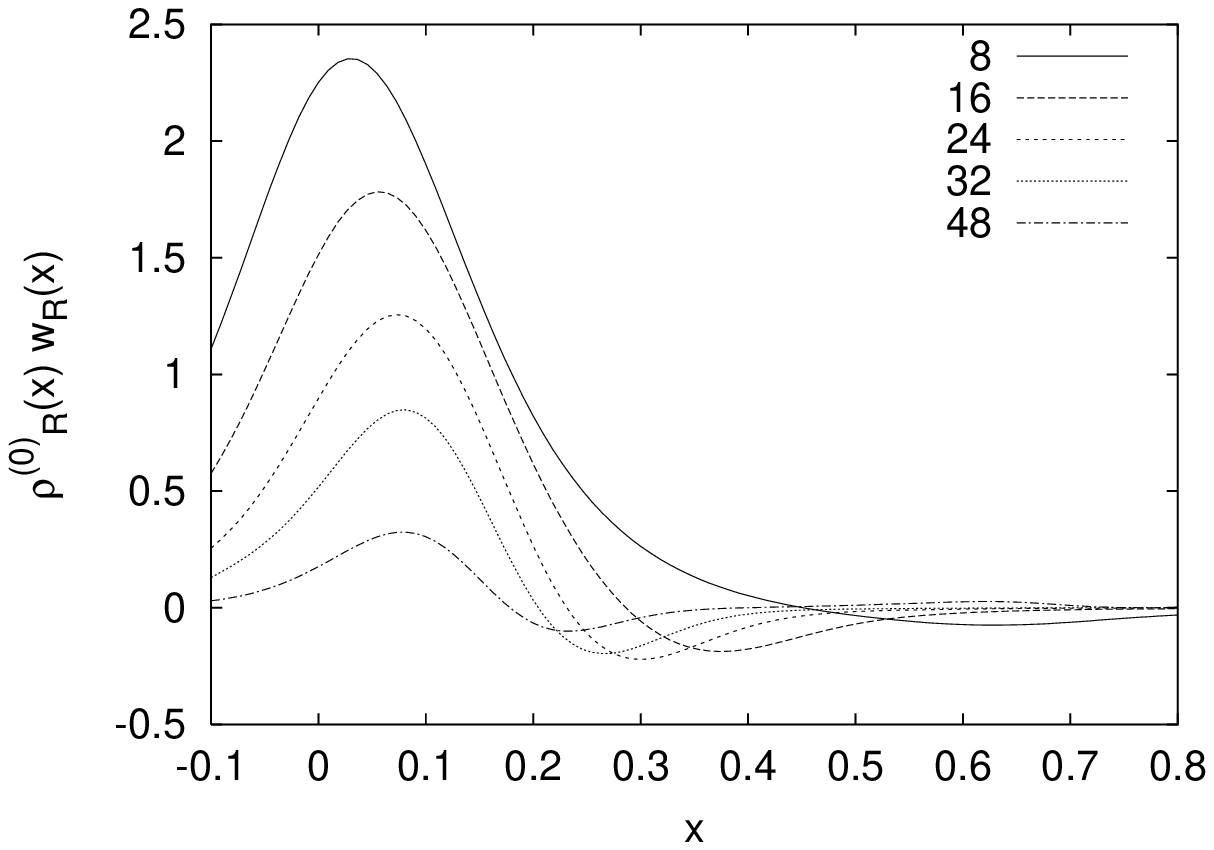,width=75mm}}\\
%\end{center}
%{\bf Fig. 2}: {\it $\rho_R(x)$ for RMT for $\mu=0.2$.}
%%%%%%%%%% END FIGURE

%%%%%%%%%% FIGURE 
%\vskip 6mm
%  \begin{center}
 {\epsfig{file=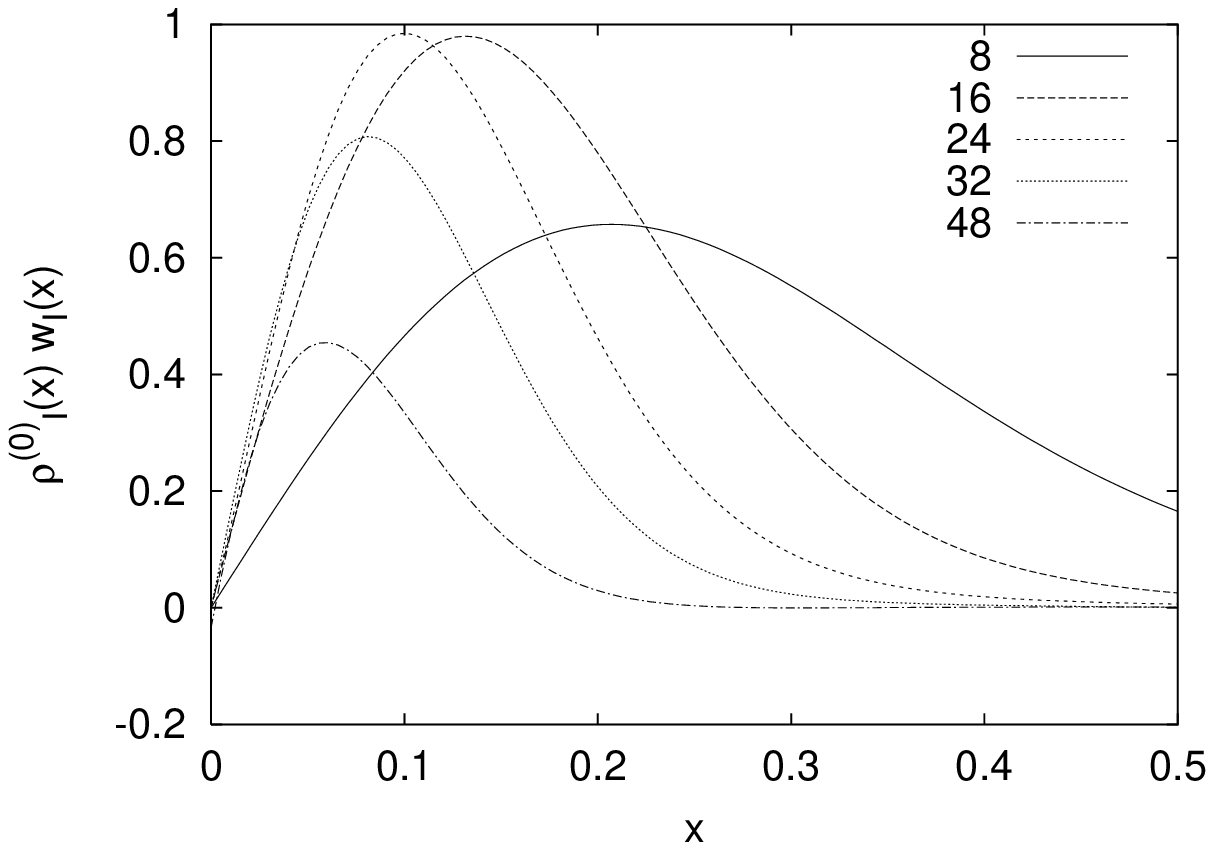,width=75mm}}
\end{center}
{\bf Fig. 2}: {\it $\rho_R(x)$ and $\rho_I(x)$ for $\mu=0.2$.}
\vskip 6mm
%%%%%%%%% END FIGURE

\section{RMT OF FINITE DENSITY QCD}

We consider RMT with one quark flavour and zero quark mass
\cite{rmt1}. The model is chosen in order to study the correctness and
effectiveness of the factorization method, since one can compare
results with known analytical solutions even for finite $N$. The
observable we measure is the ``quark number density'' $\nu$ as a
function of the chemical potential $\mu$, and we
consider the distribution functions $\rho_i(x)$, where $i=R,I$
corresponds to the real and imaginary parts of $\nu$
respectively. Notice that the effect of $\Gamma$ is dramatic, causing
a discontinuous transition in $\nu$. In
\cite{rmt} our results nicely reproduce the exact results known for
finite $N$ and we are able to achieve large enough values of $N\le 48$
to obtain the thermodynamic limit. Unfortunately, the function
$w_R(x)$ is not positive definite and the important contributions come
from the region where it changes sign. As expected, we find that
finite size scaling does not work as well as in the case of the IKKT
model (although we obtain agreement up to order of magnitude for the
values of $N\le 96$ that we explored). We also find it very difficult
to explore the crossover region near the phase transition point
$\mu_c=0.527\ldots$ for $N>8$ since $|w_i(x)|$ becomes very small. 
Since RMT is a schematic model of finite density QCD, we expect
that the factorization method will be useful to explore the phase
diagram of QCD.

%%%%%%%%% FIGURE 
\vskip 6mm
  \begin{center}
 {\epsfig{file=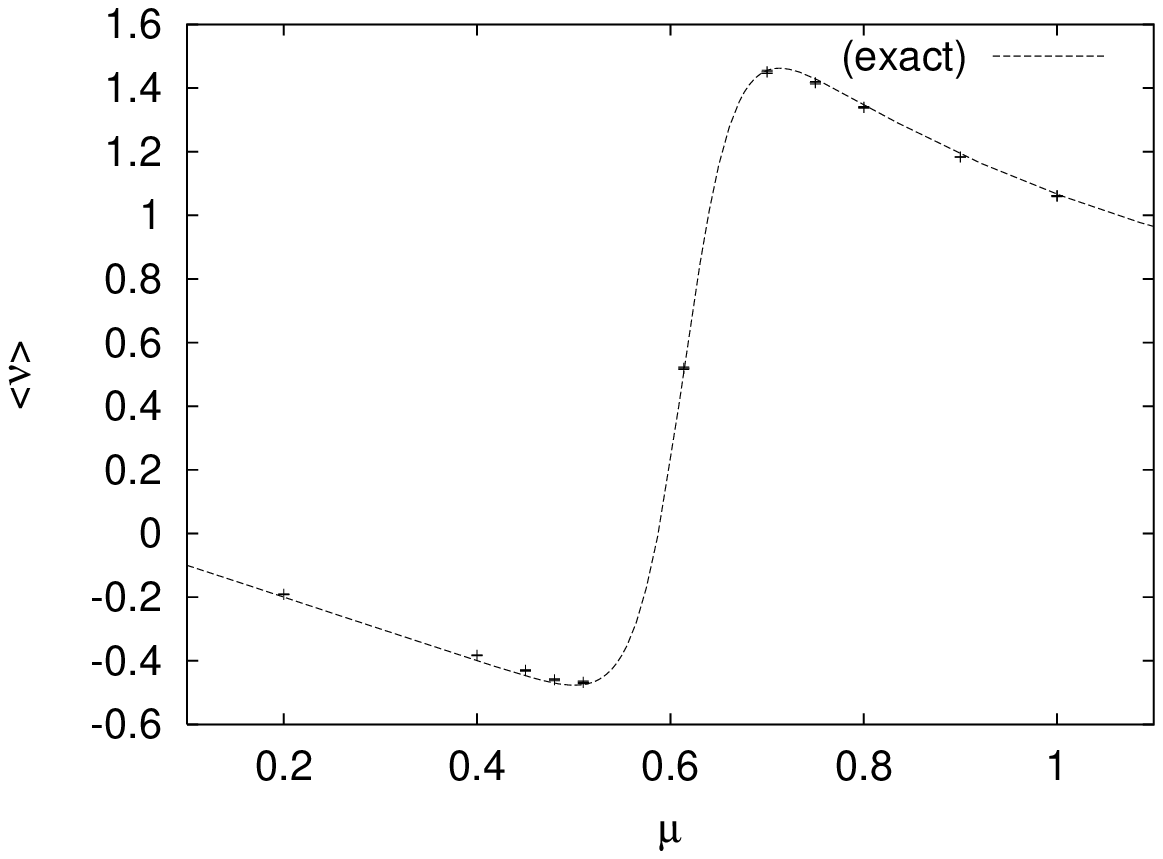,width=75mm}}
\end{center}
{\bf Fig. 3}: {\it $\vev{\nu}$ for $N=8$.}
\vskip 6mm
%%%%%%%%% END FIGURE

\end{document}